# Are the contributions of China and Korea upsetting the world system of science?

*Scientometrics* (forthcoming)

Loet Leydesdorff
Amsterdam School of Communications Research (ASCoR), University of Amsterdam
Kloveniersburgwal 48, 1012 CX  Amsterdam, The Netherlands;
loet@leydesdorff.net; http://www.leydesdorff.net
&
Ping Zhou
Information Research and Analysis Center
Institute of Scientific and Technical Information of China,
15 Fuxing Road, Beijing 100038, P. R. China; zhoup@istic.ac.cn

**Abstract**
Institutions and their aggregates are not the right units of analysis for developing a science policy with cognitive goals in view. Institutions, however, can be compared in terms of their performance with reference to their previous stages. King's (2004) 'The scientific impact of nations' has provided the data for this comparison. Evaluation of the data from this perspective along the time axis leads to completely different and hitherto overlooked conclusions: a new dynamic can be revealed which points to a group of emerging nations. These nations do not increase their contributions marginally, but their national science systems grow endogenously. In addition to publications, their citation rates keep pace with the exponential growth patterns, albeit with a delay. The center of gravity of the world system of science may be changing accordingly.

## 1. Introduction

In a recent paper in *Nature* entitled 'The scientific impact of nations,' King (2004; cf. Evidence, 2003) provided interesting data for the comparisons among nations, but in the analysis the author stumbled over the classical problem of comparing apples with oranges. Average citation rates (and to a lesser extent average publication rates) differ among fields of science and even among specialties within fields of science. For example, papers in mathematics provide fewer citations than papers in the life sciences. Impact factors of journals in immunology can be on average an order of magnitude higher than in toxicology. Thus—as a policy implication—one might conclude that closing down a country's mathematics departments would improve its citation rate. Analogously, a medical school might be advised to close down its toxicology unit in order to increase its standing on the national ranking.

In his 1985 critique of this 'institutional' approach to evaluation in terms of scientometric indicators, Collins criticized the champions of 'evaluative bibliometrics' of that time (Martin & Irvine, 1983; Narin & Carpenter, 1975) as follows:

> Irvine & Martin have studied science by breaking it up into units of comparsion defined by what I will call 'non-cognitive boundaries'. Usually the boundaries chosen have been those of institutions—the laboratory, the university department, the discipline, the nation. […] However, to develop a policy with cognitive goals in view, it is essential to start by disaggregating science according to cognitive rather than institutional boundaries—that is, to think of science as being made up of sets of research areas which involve scientists who interact, or mutually refer, *across* institutional boundaries, because of their common cognitive interests. The boundaries of such areas do not necessarily map on to the boundaries of institutions. (Collins, 1985: 554f.)

Martin & Irvine (1985) replied that one should only compare 'like with like,' that is, for example, astronomy departments in France with those in the UK. Cross-tabulation of the institutional and the cognitive dimensions may then lead to meaningful results provided that one carefully chooses the direction for the normalization. The comparison of astronomy departments across nations, for example, might tell us what both institutions have contributed to the development of astronomy worldwide, but not to the development of these units in their respective nations.

Nations can be expected to maintain a portfolio of differently positioned units in a wide range of sciences (May, 1997). However, the deployment strategies and the priorities of S&T policies can be expected to differ among nations, particularly when one compares nations in different world regions. A large number of authors in developing countries works jointly with scholars in the developed countries and can thus be assimilated in a 'world system of science' (Wagner, 2004; Wagner & Leydesdorff, 2005). However, a third group of nations have developed their scientific resources on the basis of indiginous priorities. China and Iran are obvious examples of countries which until recently have operated in relative isolation from the 'world system of science.' Korea, Taiwan, and Singapore are interesting cases because they follow a western pattern of development, but with a strong reference to China (Leydesdorff, 2003).

In summary, an evaluator should more carefully take into account the portfolio of a country. However, the delineation of fields of science in terms of scientific journal sets is not a *sine cura*. For administrative purposes policy analysts often use the categories provided by the ISI in the *Journal Citation Reports* (Pudovkin & Garfield, 2002; Small & Garfield, 1985). However, some of these categories are too wide and others too narrow.

For example, the ISI subsumes 291 journals under the heading of 'biochemistry & molecular biology,' which is far too many, while only 46 journals are categorized as 'inorganic and nuclear chemistry.' A cluster of 106 journals could be attributed to this latter category in a systematic decomposition of the matrix of journal-journal citations of the *JCR* 2003, but the number of journals classified as 'biochemistry and molecular biology journals' was always smaller than the corresponding set of the ISI (Leydesdorff, 2005). Using the aggregated journal-journal citation data provided by the *JCR*, however, the analysis of journal contents can be made much more insightful than classifying journals on the face value of the words used in journal titles (Leydesdorff, 1987).

**2. Differences among nations**

Differences among nations in terms of their research portfolios and R&D deployment strategies matter also for another reason. The ISI database has a focus on the life sciences more than on the natural sciences and mathematics, and citation lists in the latter sciences are on average shorter than in the former. The averaging effects of large numbers may compensate for these differences, but in the case of smaller countries these numeric laws may have different effects than in the case of larger ones.

In another context, Park *et al.* (2004) compared the publication and patent portfolios of South Korea and the Netherlands in 2002, precisely with the objective of evaluating the position of relatively small countries neighbouring to larger ones. These two countries are reasonably comparable in terms of their numbers of publications (Table 1). Although the respective numbers of title word occurrences are proportional to the size of the sets, these words are very different in terms of the semantic organization. This is illustrated in Figures 1 and 2 which provide so-called vector-space representations using the top hundred words in both sets for the comparison (Salton & McGill, 1983; Leydesdorff, 2004). The visualizations are based on using the algorithm of Kamada & Kawai (1989)[1] as it is available in Pajek.[2]

|  | *South Korean address* | *Dutch address* |
|---|---|---|
| Number of records in the *SCI* 2002[3] | 15,127 (2.02% World share) | 18,792 (2.51%) |
| Nr of word occurrences | 144,597 | 177,707 |
| Included in the analysis | 105 words which occur more than 160 times | 102 words which occur more than 190 times |
| Included with cosine ≥ 0.1 (pictures) | 68 words | 49 words |

**Table 1**: Comparison of Korean and Dutch shares in the *Science Citation Index* 2002 for the purpose of a semantic mapping of the cognitive dimensions.

---

[1] This algorithm represents the network as a system of springs with relaxed lengths proportional to the edge length. Nodes are iteratively repositioned to minimize the overall 'energy' of the spring system using a steepest descent procedure. The procedure is analogous to some forms of non-metric multi-dimensional scaling. A disadvantage of this model is that unconnected nodes may remain randomly positioned across the visualization. Unconnected nodes are therefore not included in the visualizations below.
[2] Pajek is freely available for academic purposes at http://vlado.fmf.uni-lj.si/pub/networks/pajek .
[3] The total number of records in the SCI 2002 is 784.458.

**Figure 1:** South-Korean set of publications covered by the *Science Citation Index 2002*: 68 most frequently occurring words relate at the level of a cosine between the word distribution vectors ≥ 0.1. (Designations added, L.)

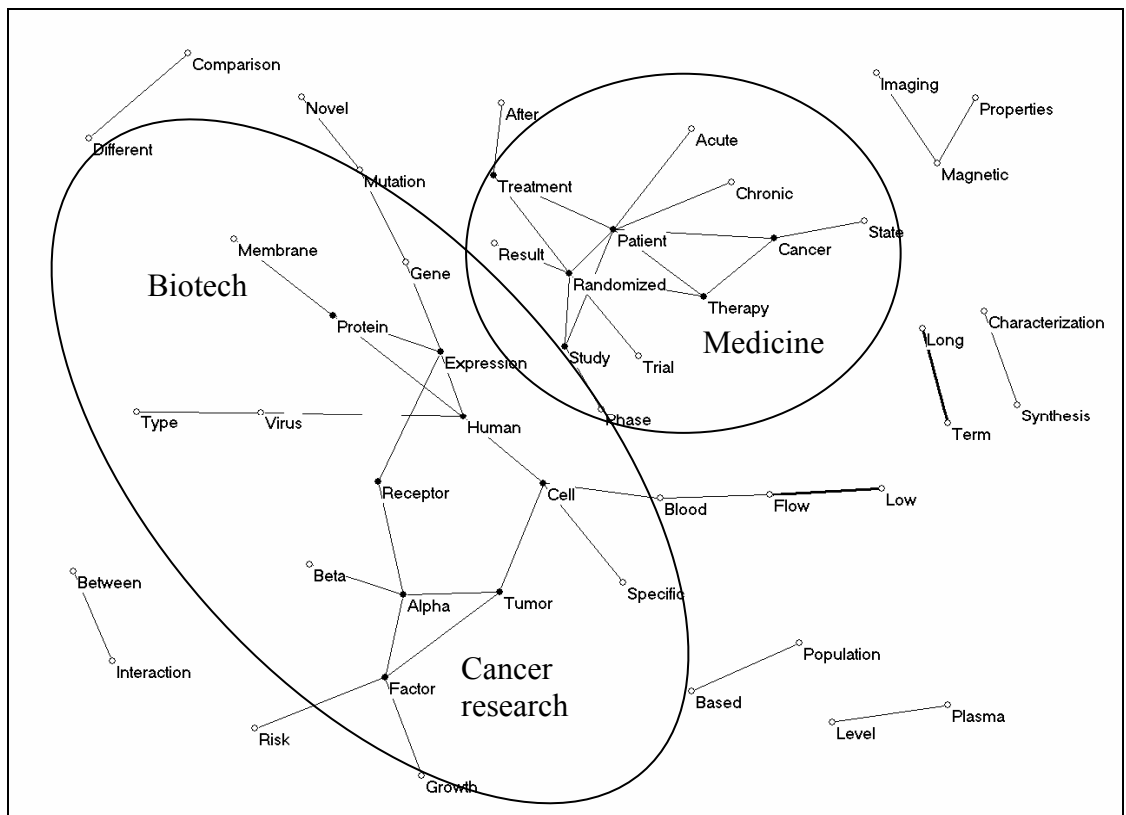

**Figure 2**: Dutch set of publications covered by the *Science Citation Index 2002*: 49 most frequently occurring words relate at the level of a cosine between the word distribution vectors $\geq 0.1$. (Designations added, L.)

Neither set is strongly organized in terms of the semantics. This accords with our point above that the sciences are not primarily organized at the national level. However, the pictures indicate the different foci in the research portfolio of these two nations. The Korean mapping provides a representation of the natural science disciplines; an otherwise unrelated group of papers focuses additionally on 'Asian medicine.' The Dutch set is concentrated in the areas of biomedicine and biotech. This focus accords with the center of the ISI database. Thus, part of the higher ranking of the Netherlands in terms of publications and citations may be due to the composition of the database. However, at the level of the database one can no longer distinguish how much of the difference is to be attributed to differences in portfolio, and how much to differences in the intrinsic quality of Korean and Dutch publications.

Unlike the exchange of currencies, the value of publications and citations is not set by an open market. The market is an equilibrating mechanism which operates at each moment in time, while the sciences develop along trajectories over time. This is acknowledged by the citation indicator: a paper may be important because it was published at a certain date. Had the paper been published later (or earlier) it might not have been cited. Nation states provide a (third) system of reference other than markets or historical trajectories. Nations can perhaps be considered as institutional arrangements that recombine very different subdynamics in a quasi-equilibrium (Aoki, 2001). Of course, it remains interesting to benchmark a nation's efforts against those of comparable units, but the more interesting comparison is then perhaps in terms of growth rates, that is, the retention of wealth from the fluxes along the time axis. The current portfolios provide us with information about the stocks, but not about the effectiveness of the portfolio.

As King noted (at p. 311), the hierarchies at the world systems level have been very stable over longer periods of time (May, 1997). Not surprisingly, therefore, the USA, the EU-15, and the UK figure at the top of the rankings in Table 1 of his paper (at p. 312). China is rated, for example, at the 20$^{th}$ position. King noted that there may be a spurious correlation between the impact and the wealth of nations. For example, China and India always appear at the lower end. Is this ranking a fair representation of the current dynamics of science and technology?

**3. The dynamics of nations**

In a reaction to King's report, Kostoff (2004) noted that when one focuses on indicators relevant for the critical field of nanotechnology, China would recently have surpassed the USA in terms of numbers of publications. Are Chinese publications so poorly cited that nevertheless one can rank China only at the twentieth place? Jin & Rousseau (2004) argued that the number of Chinese papers is increasing exponentially, but they suggested that these papers are not (yet?) of sufficient quality to cross the citation threshold.

In principle, King's tables contain answers to these questions. However, in order to see the effect of national growth and then make comparisons among nations in terms of relative growth, one has to normalize not over the columns, but within the rows of his Table 1 (at p. 312). What are the dynamics for each of the countries when the numbers for the first period (1993-1997) are compared with the numbers for the second period (1997-2001)?

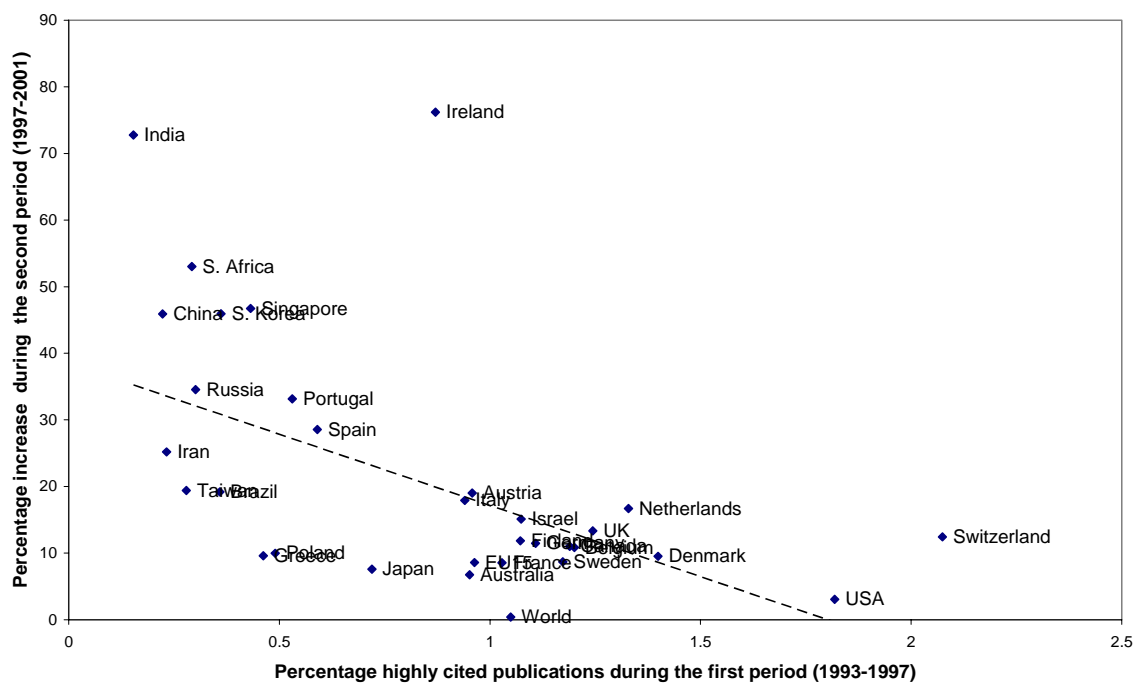

**Figure 3**: Growth of the number of most highly cited publications during the second period (1997-2001) normalized against previous contributions (1993-1997).

Let us first focus on King's prime indicator: 'the 1% most highly cited publications.' Although based on precisely the same figures, Figure 3 shows a completely different order from the one discussed in the study published in *Nature*. Among the countries showing strong growth are four nations with English as their native language: Ireland, India, South Africa, and Singapore. Chinese and South-Korean publications directly follow this group in outperforming on this indicator. Among the western countries the data for Ireland which holds the 26[th] position in King's ranking, may come as a surprise. The further increase for Switzerland should also be noted. One expects that with a higher starting value, a marginal increase will become increasingly difficult in a competitive market. (Switzerland figured, indeed, in Table 2 of King's paper in first place in the ranking.) However, let us focus on the dynamics of the emerging countries.

Are China and Korea only improving in terms of the relatively small sets of most highly cited publications? How about average publication rates? A standard indicator for this is the citation ratio per publication: $\Sigma_i\ c_i/p_i$. Figure 4 provides the normalization of this indicator in a format similar to Figure 3. The numbers for the second period are considerably lower than for the first period because an open-ended citation window was used in the study underlying King's paper (Evidence, 2003, at p. 33). (For this rather technical reason, the citation numbers are declining over time and the y-axis is negative.)

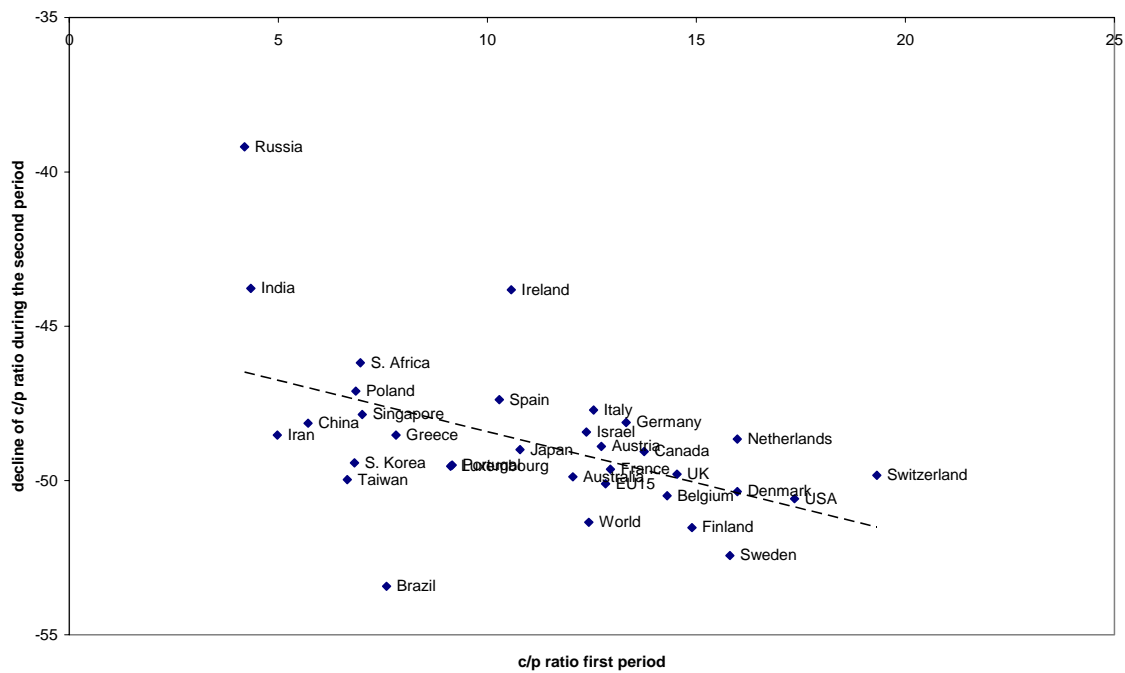

**Figure 4**: Growth of the citation/publication ratios during the second period (1997-2001) normalized against the contribution previously (1993-1997).

Figure 4 shows that Russia is the largest winner on this dimension. India, Ireland, and South Africa follow as before, but Singapore has to give way to a more average position closer to China and South Korea. Although the average citation behaviour of these countries has not improved, one should keep in mind that during this period the absolute numbers of publications from China and Korea approximately doubled. Thus, the values of the denominator (p) have changed dramatically and nevertheless the c/p ratios have been stable! On the other side, the numbers of publications with Russian and Indian addresses have decreased, as we shall see in Figure 7 below. The contributions of Russia and India can in many other respects be compared with those of European nations. Note that the European nations did not converge in terms of this indicator.

## 4. Publication and citation rates compared

*4.1 Percentages world share of publications*

Using the method of including only research articles, reviews, letters, and notes (Braun *et al.*, 1991), Figure 5 shows the percentage world share of publications for the five leading nations in science (the U.S.A., Japan, UK, Germany, and France), and the emergence of China to the sixth position during the last decade. Indeed, the trend line for China fits an exponential curve almost perfectly ($r^2 > 0.99$). The data for South Korea is also added. Although the growth rate is also spectacular, the trend line is linear in this case ($r^2 > 0.99$). This data is based on the Web-of-Science version of the *Science Citation Index* (using integer counting), but similar results for the rank-order of China in 2003 were obtained using the CD-Rom version or using fractional counting.

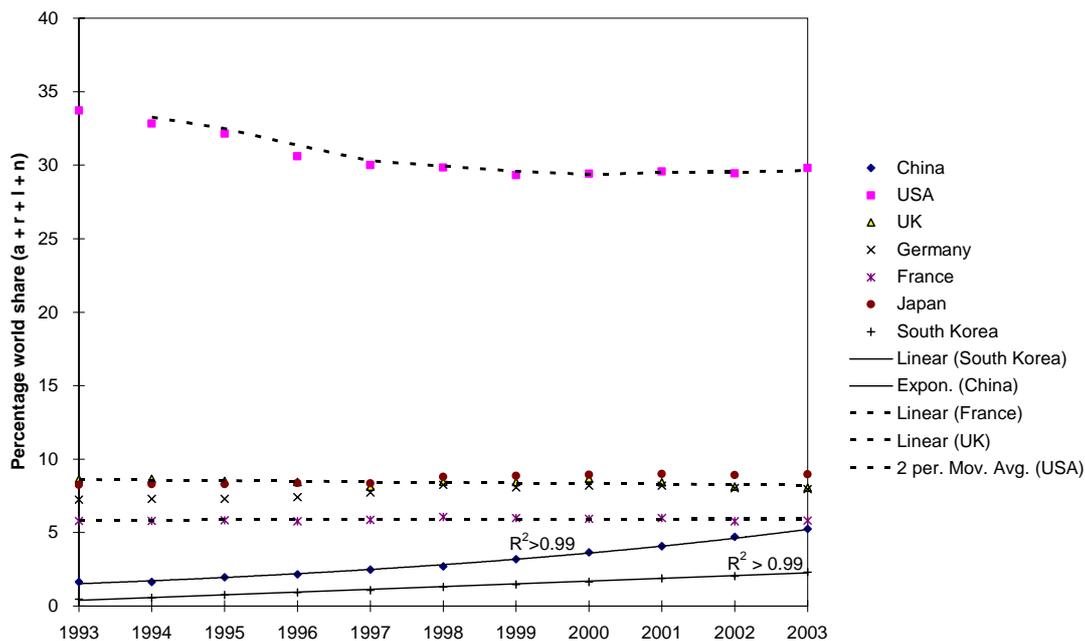

**Figure 5**: Percentage World Share of Publications for the five leading countries, China, and Korea.

In addition to the spectacular growth for China, all the major European nations and Japan have been able to increase their share of the database, but this may be an artifact of the relative decline of the USA. The data for Germany shows a strong increase in the period 1993-1998 as an effect of the unification process. Germany and the UK are since 1998 virtually of the same size in terms of this indicator. Japan, however, has surpassed the UK by obtaining the second position during the second half of the 1990s.

Papers with a UK address are much better cited than German and Japanese papers. However, this comparison of relative contributions along the column dimension of King's (2004) Table 1, that is, in terms of the world system, again obscures the underlying changes which are relative to the size of a country's contribution. A completely different order becomes visible when one compares the two periods particularly for some of the relatively smaller countries.

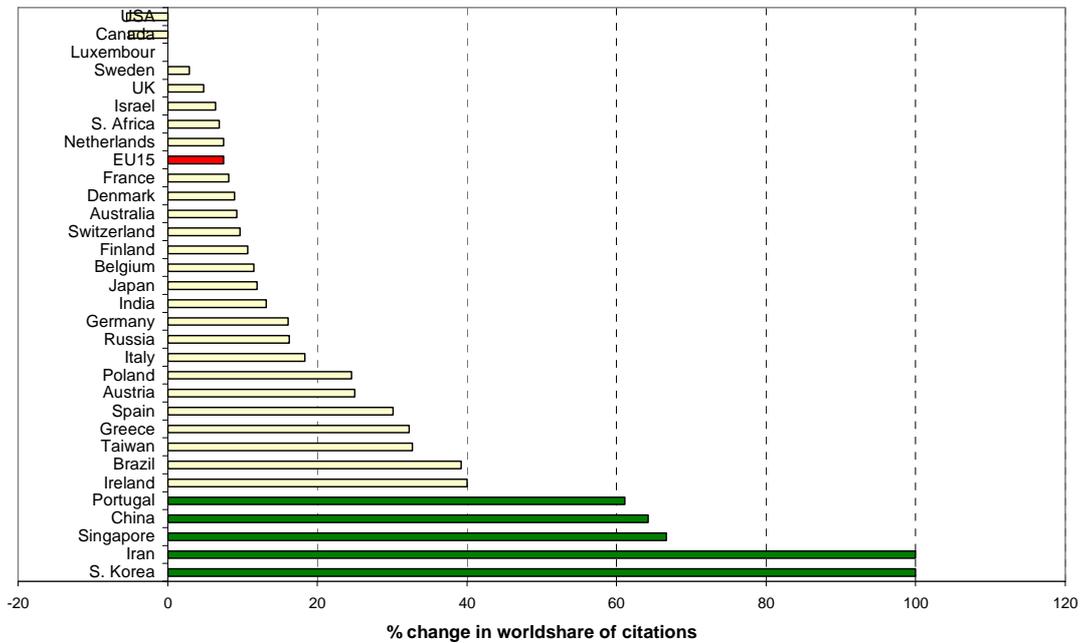

**Figure 6:** increase in the percentage of world share of citations during the period 1997-2001 when compared to the period 1993-1997 (Source: King, 2004, at p. 312)

China, Singapore, and South Korea are three of the five countries that show a spectacular increase in their citation rates when these two periods are compared at the level of individual nations. Among the European nations, Portugal is growing by more than 20 percentage points faster than Ireland. The fifth country in this league of nations is Iran. Perhapse, this is an effect of the relative opening of this nation for global exchanges during the last decade. The database contains more then ten times as many publications with an Iranian address in 2003 (2969) as in 1993 (289).

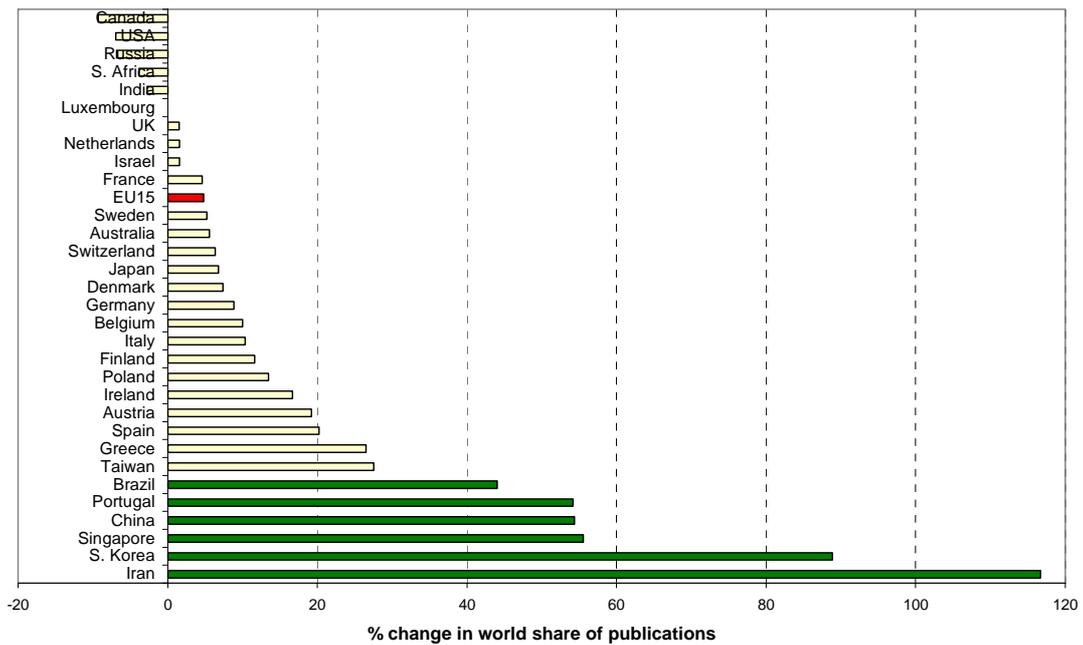

**Figure 7:** increase in the percentage of world share of publications plotted as a percentage

When the numbers of publications are analogously used as an indicator (Figure 7), Brazil joins this league of nations. The specific position of Brazil may have an effect on Portugal because these two countries share a common language. The figures also show that the increase in the c/p ratios for Russia and India were partly due to a relative decrease of the publication volumes of these two countries.

The effects for Canada on both parameters are similar to those of the USA and therefore these changes are not exclusively due to changes in the composition of the American portfolio among academic, industrial, and military research. The changes seem to reflect a genuine change of the center of gravity from North America first to Europe, but then increasingly to Asia. Note that Japan and Taiwan are not participating in this latter pattern. At the global level the rise of China is perhaps the main effect because of the volume. Singapore follows the exponential growth pattern of China, as does Iran, but the historical dynamics are probably rather different in the latter case (Figure 8).

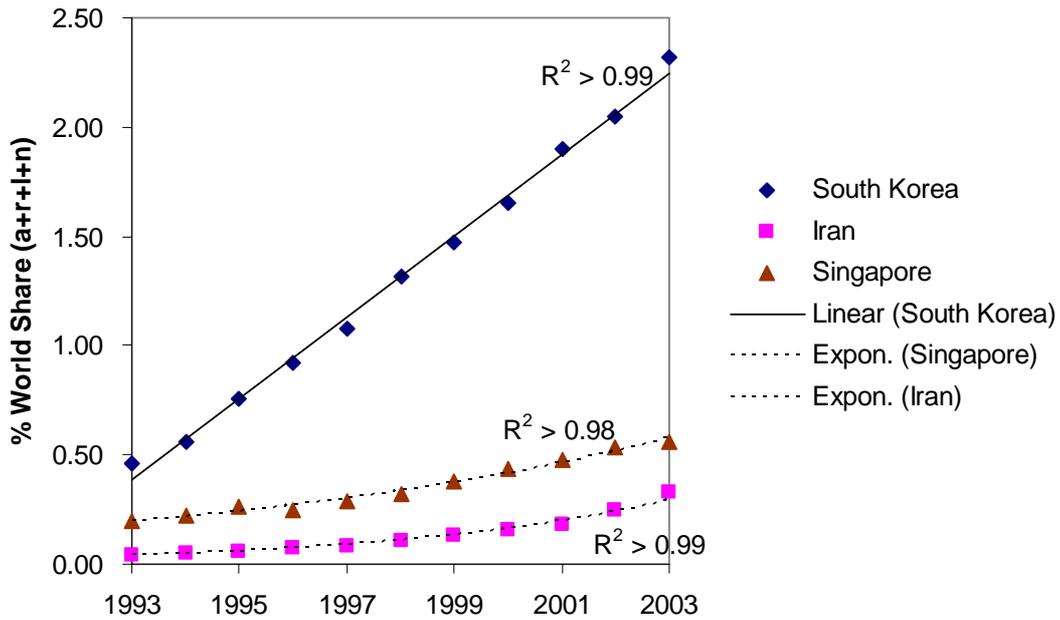

**Figure 8**: Growth of the percentage world share of publications for some fastly-growing Asian countries.

In a series of publications, the Institute of Scientific and Technical Information of China provides information about the citations to papers with a Chinese address using a citation window of ten years. This data, however, include the set of internationally coauthored articles, reviews, and letters, but only insofar as the *first* author had a Chinese address. The plot of the data (Figure 9) shows that the exponential fit even underestimates the growth of the citations.

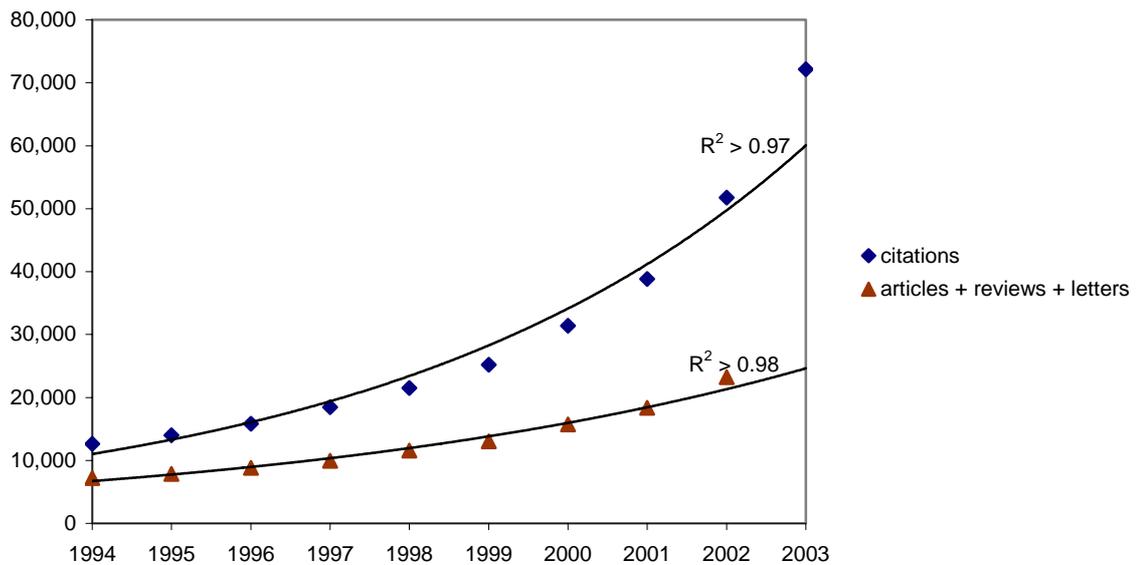

**Figure 9**: Exponential growth of the citations to papers with a Chinese address underestimates the growth pattern (Source: ISTIC, 2003 and 2004).

The most recent years exibit an enhanced growth. This may be due to the lagging of the citations on the rapidly increasing pool of Chinese publications in the international database. One may wish to argue that the number of citations is inflated because of these 'within China' citations. However, a correction for this effect would also have effects when applied to other large countries, for example, in the case of 'within USA' citations. The embeddedness of the Chinese science system in the world system through international coauthorship and citation relations merits a separate analysis (Wagner, 2004).

**Conclusions**

The previously debated decline of British science in terms of scientometric indicators (Anderson *et al*., 1988; Braun *et al*., 1991; Leydesdorff, 1989) has during the 1990s come to a definite halt. All European nations and Japan have been able to improve their performance rates gradually at the expense of the USA and Canada. However, these are marginal changes in an otherwise highly competitive balance. A completely different dynamics can be perceived at the margins of the system. Some, notably Asian, nations are able to enter the system with exponential rates of increase.

We have observed linear increase rates for the Netherlands and the Scandinavian countries during the 1980s and for Japan, Italy, and Spain during the 1990s because of ongoing tendencies to publish increasingly in English instead of their native languages. Similarly, the German contribution was affected by unification as a one-time shock (Leydesdorff, 2000). Today, however, we witness exponential growth which seems to be driven by hitherto unlimited supplies of new manpower.

From this perspective, the growth rates for China and Singapore require an explanation different from the ones for South Korea and Portugal. The latter countries can be considered as late arrivers in the group of otherwise more advanced nations. South Korea, for example, became an OECD member state in 1996, and therefore a linear growth pattern for this country could be expected. However, the size and the duration of the effect is larger than we have seen previously. The patterns of Iran and Singapore also indicate another mechanism of growth than we have hitherto seen among OECD countries. In the case of Singapore this effect can directly be related to the Chinese development, but in the case of Iran another (historical) explanation is needed. However, the size of the increasing Chinese contribution makes this disturbance of the world system of science historically unprecedented. The exponential growth rates indicate a self-reinforcing growth pattern which is possible because of a virtually unlimited reservoir of human resources with scientific competences that seems to flow into the world system with increasing speed.

**References**

Anderson, J., P. M. D. Collins, J. Irvine, P. A. Isard, B. R. Martin, and F. Narin (1988). On-Line Approaches to Measuring National Scientific Output--A Cautionary Tale. *Science and Public Policy 15*, 153-161.
Aoki, M. (2001). *Towards a Comparative Institutional Analysis*. Cambridge, MA: MIT Press.

Braun, T., W. Glänzel, & A. Schubert. (1991). The Bibliometric Assessment of UK Scientific Performance--Some Comments on Martin's Reply. *Scientometrics,* 20, 359-362.

Collins, H. M. (1985). The Possibilities of Science Policy. *Social Studies of Science,* 15, 554-558.

Evidence (2003). PSA target metrics for the UK Research Base. London: UK Office of Science and Technology, October 2003; at http://www.ost.gov.uk/policy/target_metrics.pdf (last visited on 23 October 2004).

ISTIC (2003). 2002 年度中国科技论文统计结果 (Statistics about Chinese science and technology publications in 2002). Beijing: ISTIC; Available at http://att.nst.pku.edu.cn/iy2m51xefb050hdal1zt.pdf [Retrieved on 20 December 2004].

ISTIC (2004). 2003 年度中国科技论文统计结果 (Statistics about Chinese science and technology publications in 2003). Beijing: ISTIC; Available at http://168.160.12.27/cstpcd/ViewInfoText.jsp?infoid=98340 [Retrieved on 20 December 2004].

Jin, B., & R. Rousseau. (2004). Evaluation of Research Performance and Scientometric Indicators in China. In H. F. Moed, W. Glänzel & U. Schmoch (Eds.), *Handbook of Quantitative Science and Technology Research* (pp. 497-514). Dordrecht, etc.: Kluwer Academic Publishers.

Kamada, T., and S. Kawai. 1989. An algorithm for drawing general undirected graphs. *Information Processing Letters* 31(1), 7-15.

King, D. A. (2004). The Scientific Impact of Nations. *Nature,* 430 (15 July 2004), 311-316.

Kostoff, R. (2004). The (Scientific) Wealth of Nations. *The Scientist,* 18(18), 10.

Leydesdorff, L. (1987). Various Methods for the Mapping of Science. *Scientometrics 11*, 291-320.

Leydesdorff, L. (1989). The *Science Citation Index* and the Measurement of National Performance in Terms of Numbers of Publications. *Scientometrics,* 17, 111-120.

Leydesdorff, L. (2000). Is the European Union Becoming a Single Publication System? *Scientometrics,* 47(2), 265-280.

Leydesdorff, L. (2003). The Mutual Information of University-Industry-Government Relations: An Indicator of the Triple Helix Dynamics. *Scientometrics,* 58(2), 445-467.

Leydesdorff, L. (2004). The University-Industry Knowlege Relationship: Analyzing Patents and the Science Base of Technologies. *Journal of the American Society for Information Science & Technology,* 55(11), 991-1001.

Leydesdorff, L. (2005). Can the Hierarchy among the Sciences Be Mapped in Terms of Aggregated Journal-Journal Citation Relations? *in preparation*.

Martin, B., & J. Irvine. (1983). Assessing Basic Research: Some Partial Indicators of Scientific Progress in Radio Astronomy. *Research Policy,* 12, 61-90.

Martin, B. R., & J. Irvine. (1985). Evaluating the Evaluators. *Social Studies of Science,* 15, 558-585.

May, R. M. (1997). The Scientific Wealth of Nations. *Science,* 275(5391), 793-796.

Narin, F., & M. P. Carpenter. (1975). National Publication and Citation Comparisons. *Journal of the American Society of Information Science*, 26, 80-93.

Park, H., L. Leydesdorff, H. D. Hong, & S. J. Hung. (2004). *Indicators for the Knowledge-Based Economy: A Comparison between South Korea and the*

*Netherlands.* Paper presented at the Sixth International Conference on Social Science Methodology (RC33), 17-21 August 2004, Amsterdam.

Pudovikin, A. I., & E. Garfield. (2002). Algorithmic Procedure for Finding Semantically Related Journals. *Journal of the American Society for Information Science and Technology,* 53(13), 1113-1119.

Salton, G., & M. J. McGill. (1983). *Introduction to Modern Information Retrieval.* Auckland, etc.: McGraw-Hill.

Small, H., & E. Garfield. (1985). The Geography of Science: Disciplinary and National Mappings. *Journal of Information Science,* 11, 147-159.

Wagner, C. S. (2004). *International Collaboration in Science: A New Dynamic for Knowledge Creation.* Unpublished Ph.D. Thesis, University of Amsterdam, Amsterdam.

Wagner, C. S., & L. Leydesdorff. (2005). Mapping the Network of Global Science: Comparing International Co-Authorships from 1990 to 2000. *International Journal of Technology and Globalization* (forthcoming).

Zhou, P. & L. Leydesdorff (2004). 2004 年中国的纳米技术研究成果世界领先 [China's research outcome in nanotech takes the lead in the world in 2004: The world position of China's publications], *Chinese S&T Daily*, 23 October 2004; at http://www.stdaily.com/gb/stdaily/2004-10/23/content_314274.htm [Retrieved on 20 December 2004].